\documentclass[conference]{IEEEtran}
\ifCLASSINFOpdf
   \usepackage[pdftex]{graphicx}
   \graphicspath{figs}
   \DeclareGraphicsExtensions{.pdf,.jpeg,.png}
\else
   \usepackage[dvips]{graphicx}
   \graphicspath{{eps}}
   \DeclareGraphicsExtensions{.eps}
\fi

\usepackage{hyperref}
\usepackage{multirow}
\usepackage{cite}
\usepackage[cmex10]{amsmath}

\begin{document}
\title{A Low-Power Content-Addressable-Memory Based on Clustered-Sparse-Networks}
\author{\IEEEauthorblockN{Hooman Jarollahi\IEEEauthorrefmark{1}, Vincent Gripon\IEEEauthorrefmark{2}, Naoya Onizawa\IEEEauthorrefmark{1}, and Warren J. Gross\IEEEauthorrefmark{1}}
\IEEEauthorblockA{\IEEEauthorrefmark{1}Department of Electrical and Computer Engineering, McGill University, Montreal, Quebec H3A 0E9}
\IEEEauthorblockA{\IEEEauthorrefmark{2}Electronics Department, T\'el\'ecom Bretagne, Brest, France}
Email: \{hooman.jarollahi, naoya.onizawa\} at mail.mcgill.ca, vincent.gripon at telecom-bretagne.eu, warren.gross at mcgill.ca}
\maketitle

\begin{abstract}
A low-power Content-Addressable-Memory (CAM) is introduced employing a new mechanism for associativity between the input tags and the corresponding address of the output data. The proposed architecture is based on a recently developed clustered-sparse-network using binary-weighted connections that on-average will eliminate most of the parallel comparisons performed during a search. Therefore, the dynamic energy consumption of the proposed design is significantly lower compared to that of a conventional low-power CAM design. Given an input tag, the proposed architecture computes a few possibilities for the location of the matched tag and performs the comparisons on them to locate a single valid match. A 0.13$\mu m$ CMOS technology was used for simulation purposes. The energy consumption and the search delay of the proposed design are 9.5\%, and 30.4\% of that of the conventional NAND architecture respectively with a 3.4\% higher number of transistors. 

\end{abstract}

\section{Introduction}

A Content-Addressable-Memory (CAM) is a type of memory that can be accessed using its contents rather than an explicit address. In order to access a particular content in such memories, a search data word is compared against previously stored entries in parallel to find a match. Each stored entry is associated with a tag that is used in the comparison process. Once a search data word is applied to the input of a CAM, the matching data word is retrieved within a single clock cycle if it exists. This prominent feature makes CAM a promising candidate for applications where frequent and fast look-up operations are required such as in translation look-aside buffers (TLBs) \cite{Intel2011} and network routers  \cite{routers2}. 

Due to the frequent and parallel search operations, CAMs consume a significant amount of energy. CAM architectures typically use highly capacitive searchlines causing them not to be energy-efficient when scaled. For example, this power inefficiency has constrained TLBs to be limited to no more than 512 entries in current processors.  
Energy saving opportunities have been discovered by employing either circuit-level techniques \cite{Naoya2012}
, architectural-level \cite{precomputation, Hsieh2008} techniques or the co-design of the two, \cite{Hwang2011} some of which have been surveyed in \cite{Sheikholeslami2006}. Although dynamic CMOS circuit techniques can result in low-power and low-cost CAMs, these designs can suffer from low noise-margins, charge sharing and other problems \cite{precomputation}. 

A new family of associative memories based on clustered-neural-networks has been recently introduced \cite{Gripon2011b}\cite{Gripon2012}, and implemented using FPGAs \cite{Jarollahi2012}. Such memories make it possible to store many short messages instead of few long ones as in the conventional Hopfield neural networks \cite{Hopfield1982} with significantly lower level of computational complexity. 
In this paper, a variation of this approach and a corresponding architecture are introduced to construct a classifier that can be trained with the association between the input tags and the corresponding addresses of the output data. The proposed architecture will eliminate most of the parallel comparisons required during the search that consume a large amount of energy. The concept of this classifier is similar to that of the pre-computation based CAM (PB-CAM) \cite{precomputation, Hsieh2008}. A drawback of such methods, unlike the proposed architecture, is that as the length of the tags is increased, the delay and the circuit complexity of the precomputation stage is dramatically increased. Furthermore, we will show that unlike the PB-CAMs, the proposed architecture can potentially narrow down the search procedure to only two comparisons with a simple computational complexity.

\begin{figure}[tb]
\centering
\includegraphics[width=3.0in]{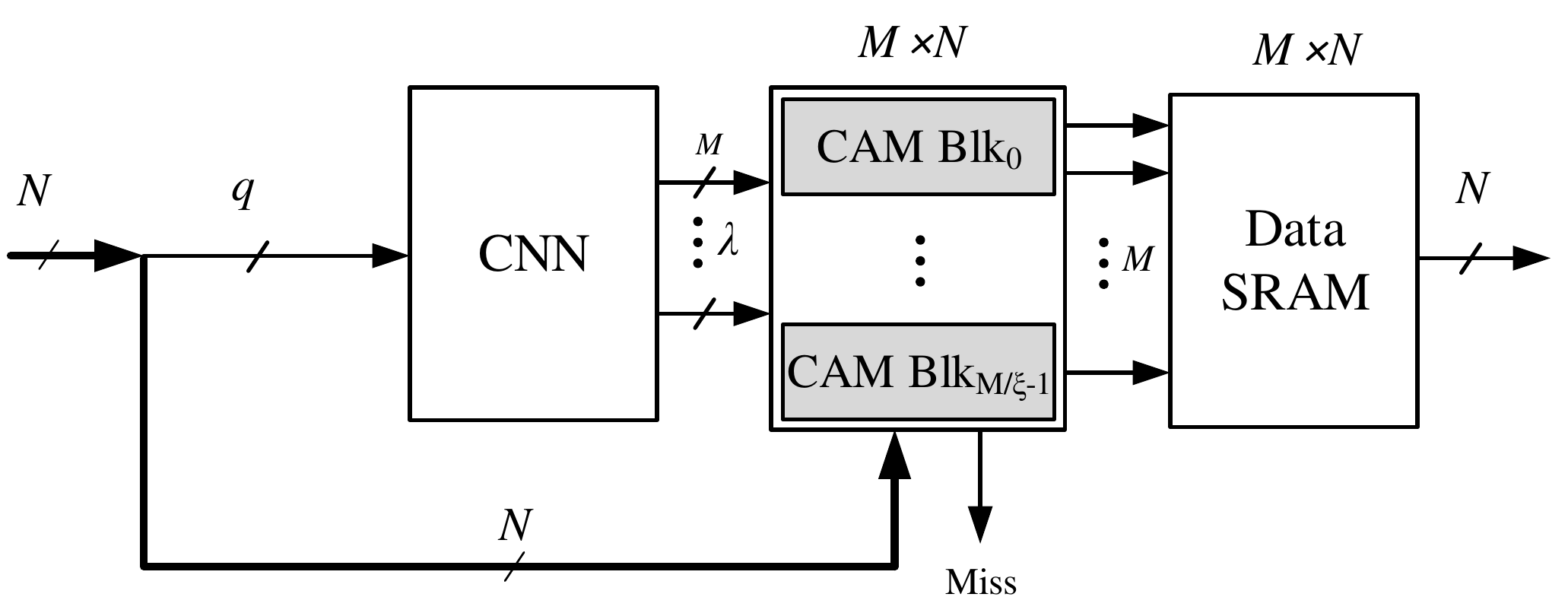}
\vspace{-3mm}
\caption{Top level block diagram of the proposed architecture. The CAM array is divided into $M/\zeta-1$ sub-blocks that can be independently activated for comparison. The compare enable signals are generated by the CNN.}

\label{fig_top}
\vspace{-5mm}
\end{figure}

The proposed architecture consists of a neural-network-based classifier coupled to a CAM array. The CAM array is divided into several equally-sized sub-blocks which can be activated independently. For a previously trained network and given an input tag, the classifier only uses a small portion of it and will predict, on average, only two out of several sub-blocks of the CAM to be activated. If the number of sub-blocks is equal to the number of entries in the CAM, only two CAM entries should be compared to find the match with the cost of higher hardware complexity. Once the sub-blocks are activated, the tag is compared against the few entries in them while keeping the rest deactivated. The total number of sub-blocks can be designed depending on the silicon area availability since each sub-block will slightly increase the silicon area. If the input data word is not uniformly distributed, more sub-blocks will be activated during a search and the accuracy of the final output is not affected. However, since the full-length of the tag is not used in the proposed architecture, it is possible to select the reduced-length tag bits depending on the application and according to a pattern to reduce the tag correlation. In Section \ref{sec_mech} the proposed associativity mechanism is introduced. Section \ref{sec_circuit_implementation} describes the hardware architecture followed by Section \ref{sec_exp} with the simulation results. Circuit level simulations throughout this paper are obtained using SPECTRE, in a 0.13 $\mu m$ CMOS technology. Concluding remarks are given in Section \ref{sec_concl}. 

\section{Proposed Associativity Mechanism}
\label{sec_mech}

As shown in Fig. \ref{fig_top} the proposed architecture consists of a clustered-neural-network (CNN) connected to a modified CAM array. The CNN is at first trained with the association between a reduced-length tag and the address of the data to be later retrieved. The CAM array is based on a conventional architecture but is divided into several sub-blocks that can be compare-enabled independently. Once an input tag is applied to the CNN, it will predict which CAM sub-block(s) need to be compare-enabled and thus saves power by disabling the rest. If the full length of the tag is used, the classifier will be able to always point to a single sub-block. However, training the network with the full length of the tags will affect the hardware complexity of the CNN. 
If the reduced-length tags are uniformly distributed, on average, only two possibilities are found with the right number of bits of the reduced-length tag. On the other hand, in some cases, this truncation may cause ambiguities in finding the valid match causing more than one possible CAM sub-block to be activated. This effect will not affect the accuracy of the final result but will cost more power. 

\subsection{Clustered Neural Networks (CNN)}
\label{cnn}

\begin{figure}[tb]
\centering
\includegraphics[width=3.0in]{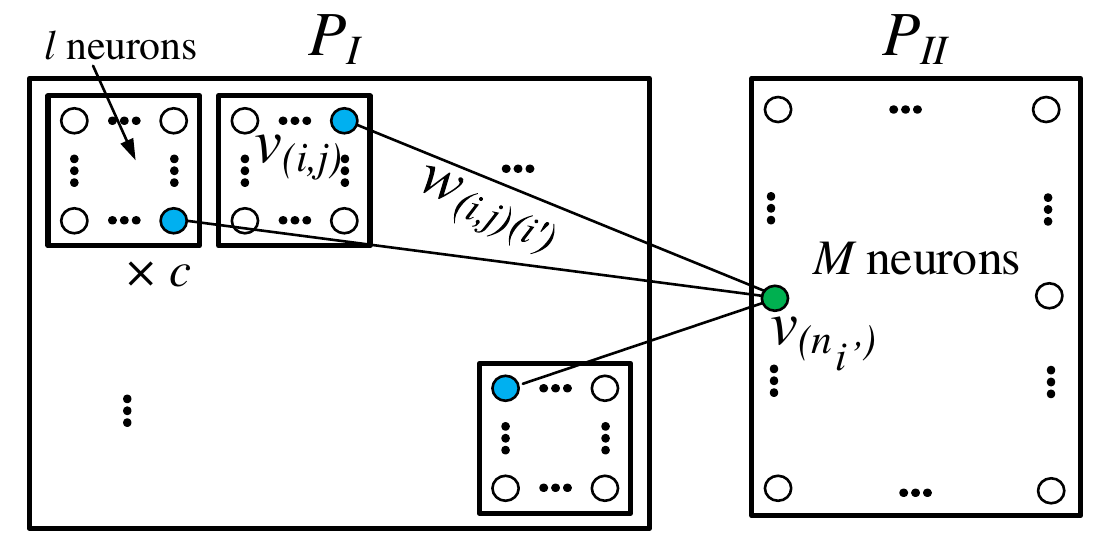}
\caption{Representation of the proposed CNN for a CAM consisting of M entries and a reduced-length tag of $c\times \text{log}_2(l)$.}
\label{fig_arch1_abstract}
\vspace{-5mm}
\end{figure}

As shown in Fig. \ref{fig_arch1_abstract}, the network consists of two parts: $P_I$ and $P_{II}$. $P_I$ corresponds to the input tag and consists of neurons that are grouped into $c$ equally-sized clusters with $l$ neurons each. Each neural value is binary, i.e. it is either activated or not. The processing of an input message can be within either of the two situations: training or decoding. 
In this paper, either for training or decoding purposes, the input tag is reduced in length to $q$ bits, and then divided into $c$ equally-sized partitions of length $\kappa$ bits each. Each partition is then mapped into a neuron in its corresponding cluster using a direct binary-to-integer mapping from the tag portion to the index of the neuron to be activated. Thus $l=2^\kappa$. If $l$ is a given parameter, the number of clusters is calculated to be $c=q/\log_2(l)$. It is important to note that there are no connections within the neurons and clusters inside $P_I$. 
 
$P_{II}$ is a single cluster consisting of $M$ neurons which is equal to the number of entries in the CAM. Each neuron in $P_{II}$, $n_{i'}$, is connected to every neuron in $P_I$ holding a binary weight, $w_{(i,j)(i')}$, which is either `0' or `1'. $w_{(i,j)(i')}$ indicates the weight of the j'th neuron in the i'th cluster of $P_{I}$ with the value $v_{(i,j)}$ to the neuron $i'$ of $P_{II}$ with the value $v_{n_{i'}}$. 

\subsubsection{Network Training}
The value of the connection weights are set during the training process and are stored in a memory module such that they can later be used to retrieve the address of the target data. A connection has a value `1' when there exists an association between the corresponding neuron in $P_{I}$ and a pointer to a CAM entry which is represented as a neuron in $P_{II}$. For example, if $c=2$ and $q=6$, for a truncated input tag `101110' corresponding to the fourth entry in the CAM, $w_{(1,5)(4)}$, and $w_{(2,6)(4)}$ will be equal to `1'.

\subsubsection{Tag Decoding}
Once the network has been trained, the ultimate goal after receiving the tag is to determine which neuron(s) in $P_{II}$ should be activated based on the given $q$ bits of the tag. This process is called \emph{decoding} where the weights are recalled from the memory. The decoding process is divided into four steps: I) An input tag is reduced in length to $q$ bits and divided into $c$ equally sized partitions. The $q$ bits can be selected within the tag bits in such a way to reduce the correlation. II) \emph{Local Decoding} (LD): A single neuron per cluster in $P_{I}$ is activated using a direct binary-to-integer mapping from the tag portion to the index of the neuron to be activated. III) \emph{Global Decoding} (GD): GD determines which neuron(s) in $P_{II}$ must be activated based on the results from LD and the stored weights. If there is at least one active connection from each cluster in $P_{I}$ towards a neuron in $P_{II}$, that neuron will be activated.  GD can be shown mathematically by using (\ref{equ_gd}). 

\vspace{-8mm}
\begin{eqnarray}
v_{n_{i'}} = \bigwedge_{i=1}^{c} \bigvee_{j=1}^{l}(w_{(i,j)(i')} \bigwedge v_{(i,j)})
\label{equ_gd}
\end{eqnarray}
\vspace{-2mm}

\noindent where $\bigvee$ and $\bigwedge$ represent the logic OR and AND respectively. IV) Because we may not afford in terms of area to have only one independently-controlled CAM entry per neuron in $P_{II}$, the neurons in $P_{II}$ are grouped into $\zeta$-neuron partitions. A logic OR is performed on the neural values of each group. Therefore the final outputs $v_{n_i}$ will form $M/\zeta$ bits which will enable parallel comparisons in the corresponding CAM sub-blocks as shown in Fig. \ref{fig_CAM_array}. 

\subsection{Tag Length Reduction}
Given the input tags, the number of bits in the reduced-length tag, $q$, will determine the number of ambiguities in $P_{II}$. These ambiguities will require additional comparisons to find the exact match in the CAM. On the other hand, no truncation means no ambiguities but a higher level of hardware complexity in the CNN. Therefore, even if the full-length tags are not uniformly distributed, depending on the application it is possible to select the bits in the reduced length tag in such a way to reduce correlations.

\begin{figure}[tb]
\centering
\includegraphics[width=2.9in]{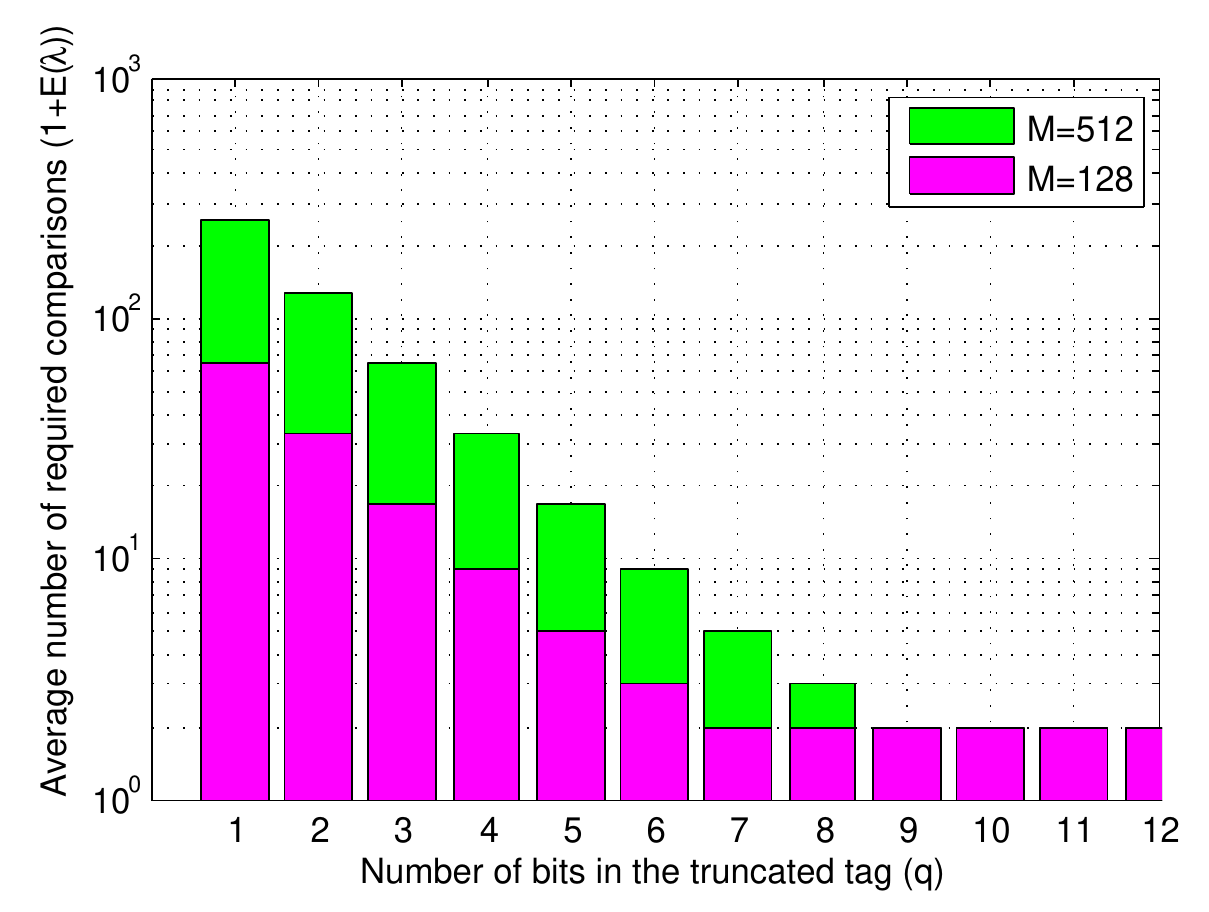}
\caption{Expected value of the number of required comparisons in the proposed architecture vs. the number of bits in the reduced-length tag.}
\label{fig_avg_comparisons_M_512_M_128_q}
\vspace{-2mm}
\end{figure}

Fig. \ref{fig_avg_comparisons_M_512_M_128_q} depicts simulation results based on one million uniformly-random reduced-length tags and two different CAM sizes. It shows how the expected value of the number of ambiguities ($E(\lambda)$) is decreased to only one by increasing the value of the number of bits in the reduced-length tag.

\section{Circuit Implementation}
\label{sec_circuit_implementation}
A top-level block-diagram of the implementation of the proposed architecture is depicted in Fig. \ref{fig_top}. It shows how the CNN module is connected to a custom designed CAM module depicted in Fig. \ref{fig_CAM_array}. A 9-transistor (9T) XOR-type CAM with NOR-type Matchline (ML) architecture was used for simulation purposes. Conventional NAND and NOR-type CAM architectures were also implemented for comparison purposes.

In order to implement a circuit that can elaborate the benefit of the proposed algorithm, a set of design points were selected among 15 different parameter sets with the common goal of discovering the minimum energy consumption per search, while keeping the silicon area overhead and the delay reasonable. The optimum design choices based on the experimental simulations for 512 CAM entries are summarized in Table \ref{table_design_param}.

\begin{table}[!tb]
\caption{Reference Design Parameters}
\begin{center}
\begin{tabular}{c|c|c}
&Parameter & Value\\ \hline\hline

	&$M$          & 512  \\ \cline{2-3}
	&$N$          & 128   \\ \cline{2-3}
	&$\zeta$      & 8    \\ \cline{2-3}
	&$\beta$      & 64   \\ \cline{2-3}
CNN	&$E(\lambda)$ & 1    \\ \cline{2-3} 
	&$q$          & 9    \\ \cline{2-3}
	&$c$          & 3    \\ \cline{2-3}
	&$l$          & 8    \\ \hline\hline
	&CAM type     & XOR \\ \cline{2-3} 
CAM	&ML Arch.     & NOR \\ \hline\hline
	&Supply Voltage & 1.2V \\ \cline{2-3} 
	&Technology     & 0.13 $\mu m$ \\

\multicolumn{2}{c}{…}
\end{tabular}
\end{center}
\vspace{-10 mm}
\label{table_design_param}
\end{table}

\subsection{CNN Architecture}
\begin{figure}[tb]
\centering
\includegraphics[width=3.6in]{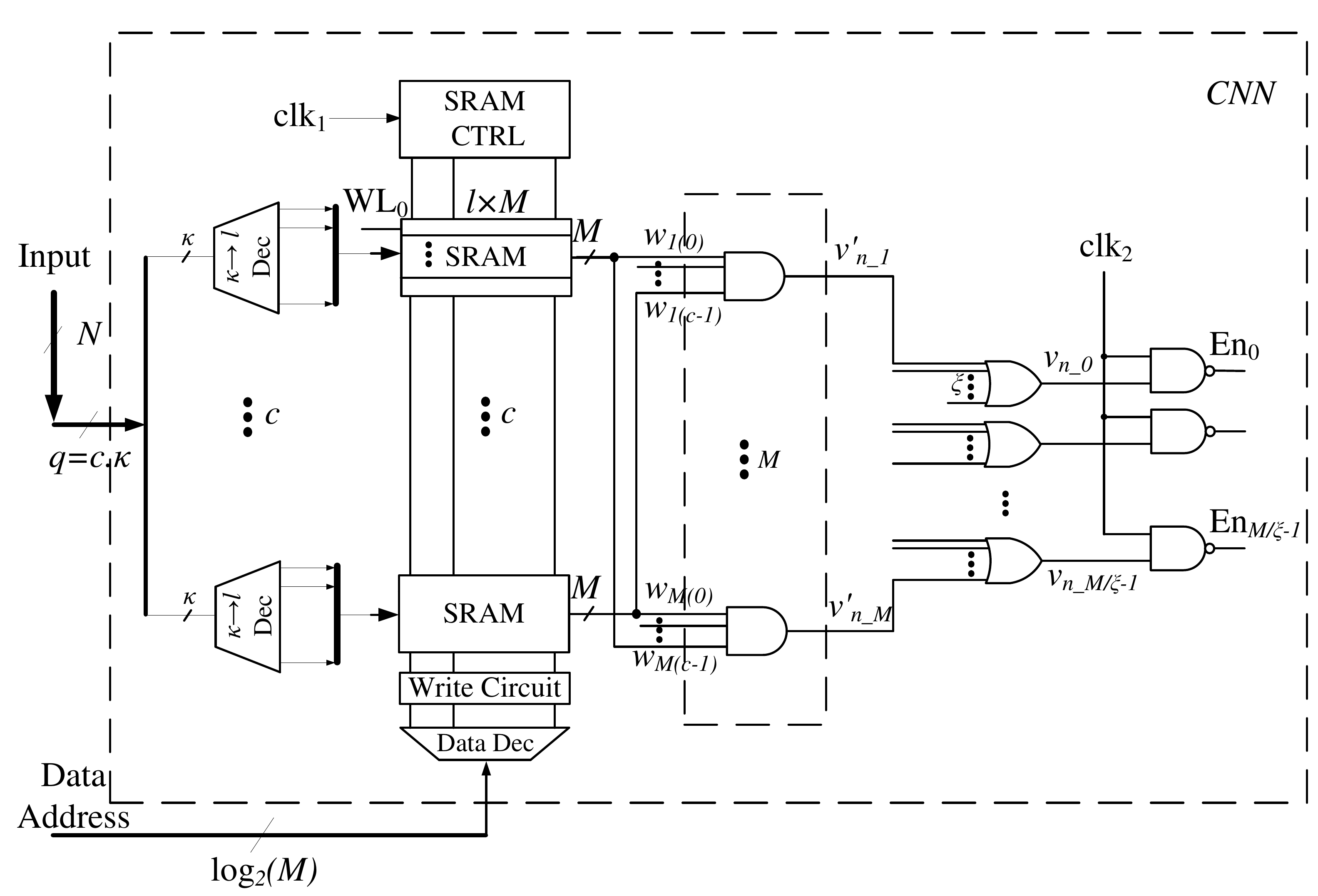}
\caption{Simplified block diagram of the proposed architecture for the CNN generating the compare-enable signals for the CAM array.}
\label{fig_CNN}
\vspace{-5mm}
\end{figure}

The architecture of the CNN is depicted in Fig. \ref{fig_CNN}. The CNN consists of $\kappa$-to-$l$ one-hot decoders, SRAM modules, and some logic elements to prepare the compare-enable signals for the CAM sub-blocks. In this architecture of the CNN, instead of calculating the neural values in $P_{II}$ by computing all of the AND and OR operations in (\ref{equ_gd}), only those connections coming from the activated neurons in $P_{I}$ are used for computation of the value of a neuron in $P_{II}$ instead of considering every connection for all neurons in $P_{I}$. This is possible by integrating the decoder and the SRAM module as shown in Fig. \ref{fig_CNN}. 

The SRAM modules are arranged into $c$ blocks with $l$ rows and $M$ columns each. Each of these blocks stores the connection weights obtained in the training process for each cluster where the association between address of the data and the reduced-length tags is stored. 

The decoding process shown in (\ref{equ_gd}) is implemented using the structure of the SRAM modules and the $c$-input AND gates. Once the input tag is reduced in length, the one-hot decoders will determine which row of the SRAM is to be accessed. In each SRAM module, this row is the only row that  can contain the information leading to the activation of a neuron in $P_{II}$ and inherently eliminates unnecessary $w_{(i,j)(i')} \bigwedge v_{(i,j)}$ operations between the weights and the neural values in (\ref{equ_gd}).

\begin{table*}[t]
\renewcommand{\arraystretch}{0.8}
\begin{center}
\caption{Result Comparisons.}
\vspace{0mm}
\begin{tabular}{c||c|c|c|c|c|c|c}
\hline
&               PF-CDPD        & Hybrid           & STOS            & HS-WA           & Ref. NAND      & Ref. NOR      & Proposed      \\
&               \cite{AND1}    & \cite{Liao2008}  &\cite{Naoya2012} & \cite{Intel2011}&                 &                &               \\
\hline
\hline
Configuration  & 256$\times$128& 128$\times$32    & 256$\times$144  & 128$\times$128  & 512$\times$128  & 512$\times$128 & 512$\times$128 \\
\hline
CAM type       & BCAM          & BCAM             & BCAM            & BCAM            & BCAM            & BCAM           & BCAM \\
\hline
Cell type      & NAND          & NAND-NOR         & NAND            & NAND-NOR        & NAND            & NOR            & NOR  \\
\hline
Technology     & 0.18 $\mu m$  & 0.13 $\mu m$     & 90 $nm$         & 32 $nm$         & 0.13 $\mu m$    & 0.13 $\mu m$   & 0.13 $\mu m$\\
\hline
Delay [ns]     & 2.10          & 0.60             & 0.26            & 0.145           & 2.30             & 0.55          & 0.70  \\
\hline
Energy metric [fJ/bit/search]  & 2.33 & 1.3       & 0.162           & 1.07            & 1.30            & 2.39          & 0.124 \\
\hline
\end{tabular}
\vspace{-5mm}
\label{table_performance}
\end{center}
\end{table*}

\subsection{CAM Architecture}
\begin{figure}[tb]
\centering
\includegraphics[width=2.8in]{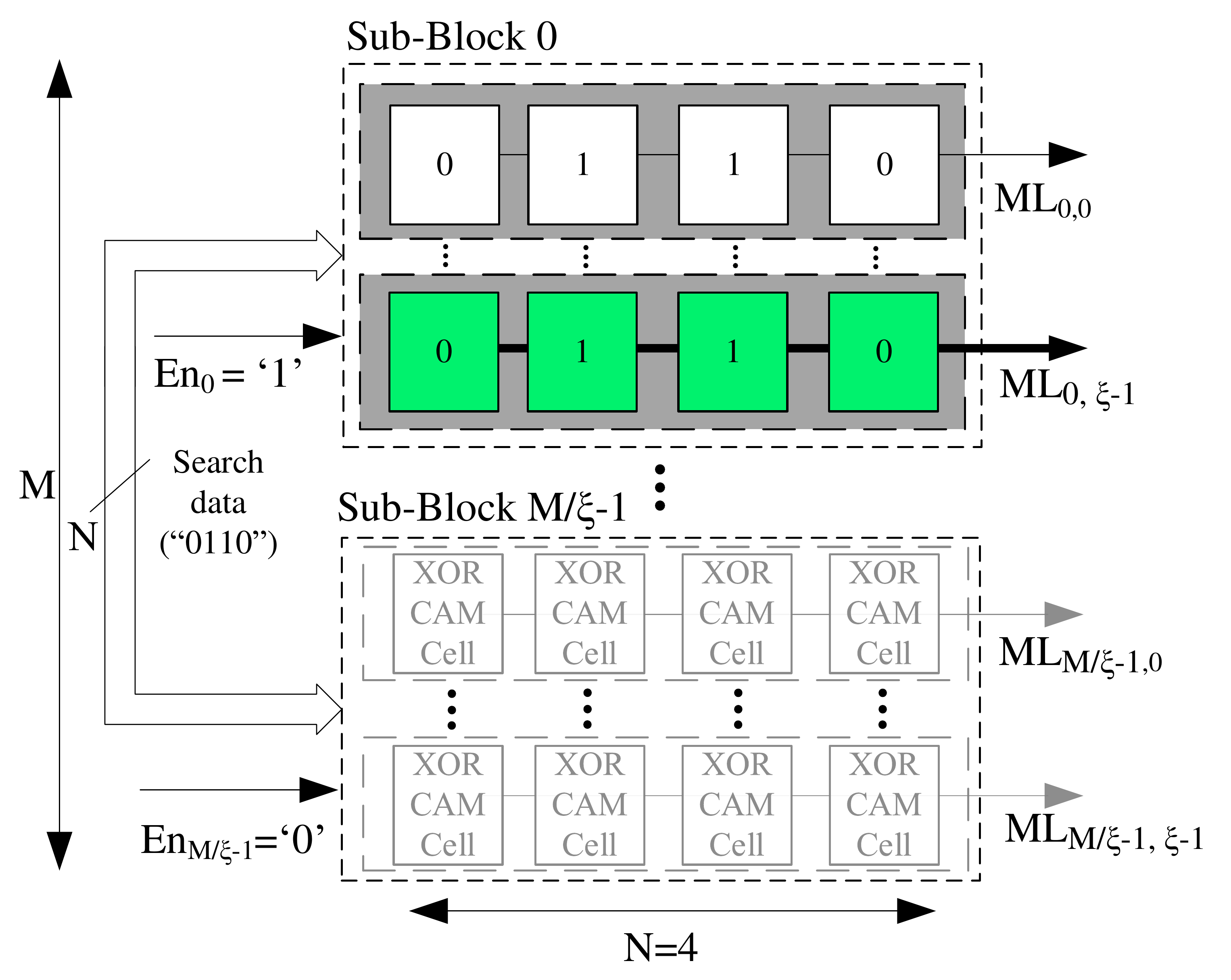}
\vspace{-4mm}
\caption{Simplified array organization of the proposed CAM architecture showing an example when $N=4$, Search data word is $``0110"$ and $En_0$=`1'. The sub-block compare-enable signals are generated by the CNN.}
\label{fig_CAM_array}
\vspace{-5mm}
\end{figure}

In order to exploit the prominent feature of the CNN in classification of the search data, the conventional CAM needs to be divided into sufficient number of compare-enabled sub-blocks such that 1) The number of sub-blocks should not be too many to expand the layout and to complicate the interconnections 2) The number of sub-blocks should not be too few to be able to exploit to energy-saving opportunity with the CNN. Consequently, the neurons in $P_{II}$ are grouped and ORed as shown in Fig. \ref{fig_CNN} to construct a compare-enable signal for a CAM sub-block.  The number of sub-blocks, $\beta$, is equal to $M/\zeta$, where $M$ is the total number of entries of the CAM, and $\zeta$ is the number of CAM rows per sub-block in the hierarchical arrangement as shown in Fig. \ref{fig_CAM_array}.  

Since optimally only two sub-block are activated on average using the proposed architecture, it is possible to exploit the low-latency feature of the NOR architecture in the proposed design. This way, we will still reduce the energy consumption compared to that of the conventional NAND architecture. 

\section{Experimental Results}
\label{sec_exp}
A wave-pipelining approach under worst-case process  conditions (slow-slow) has been followed for $clk_1$ and $clk_2$ signals in Fig. \ref{fig_CNN} to integrate the CNN and the CAM module. Other methods such as registered pipelining are also possible. The number of transistors in the proposed design is 3.4\% higher than that of the conventional design.
In the simulations for measuring the energy consumption per search half of the data bits were assumed to mismatch in case of a word mismatch. 

The delay is measured by the maximum reliable frequency of operation in the worst-case delay scenario. Table \ref{table_performance} shows the comparison of the proposed architecture with some other works including our own simulations for conventional NAND and NOR CAMs. The energy consumption and the delay of the proposed design can be converted to 90 $nm$ CMOS technology as in \cite{Naoya2012} ($V_{DD} = 1.0$ V) for comparison purposes using the method in \cite{Hwang2011}. The projected values are equal to 0.060 fJ/bit/search and 0.582 $ns$ respectively. 

\section{Conclusion}
\label{sec_concl} 
In this paper, a low-power Content Addressable Memory (CAM) is introduced. The proposed architecture employs a novel associativity mechanism based on a recently developed family of neural-network-based associative memories. This architecture is suitable for low-power applications where frequent and parallel look-up operations are required. The proposed architecture employs a clustered-neural-network module which is connected to several independently-compare-enabled CAM sub-blocks. With optimized lengths of the reduced-length tags, the network will eliminate most of the comparisons given a uniformly random distribution of the reduced-length inputs. Non-uniformity will cost power but will not affect accuracy. Conventional NAND and NOR-type architectures were also implemented for comparison purposes. It has been estimated that for a selected sample design parameter for the proposed architecture, the energy consumption and the search delay of the proposed design are 9.5\%, and 30.4\% of that of the conventional NAND architecture respectively with a 3.4\% higher number of transistors. 
\bibliographystyle{IEEEtran}
\bibliography{hooman}
\end{document}